\documentclass[a4paper,fleqn,usenatbib]{mnras}

\usepackage{newtxtext,newtxmath}

\usepackage{upgreek}

\usepackage[T1]{fontenc}
\usepackage{ae,aecompl}

\usepackage{upgreek}

\usepackage{color}

\usepackage{array,booktabs}

\usepackage{graphicx}	
\usepackage{amsmath}	
\usepackage{amssymb}	
\usepackage[switch, modulo]{lineno}

\title[Probing FRB VHE \& optical emission with MAGIC]
{Constraining very-high-energy and optical emission from FRB~121102 with the MAGIC telescopes}

\author[MAGIC Collaboration]{
\parbox{\textwidth}{
MAGIC Collaboration:
V.~A.~Acciari$^{1}$,
S.~Ansoldi$^{2,20}$,
L.~A.~Antonelli$^{3}$,
A.~Arbet Engels$^{4}$,
C.~Arcaro$^{5}$,
D.~Baack$^{6}$,
A.~Babi\'c$^{7}$,
B.~Banerjee$^{8}$,
P.~Bangale$^{9}$,
U.~Barres de Almeida$^{9,10}$
% ...
\\
(Affiliations can be found after the references)
}}

\date{Accepted XXX. Received YYY; in original form ZZZ}

\pubyear{2018}

\begin{document}
\label{firstpage}
\pagerange{\pageref{firstpage}--\pageref{lastpage}}
\maketitle

\clearpage
\begin{abstract}

Fast radio bursts (FRBs) are bright flashes observed typically at GHz frequencies with millisecond duration, whose origin is likely extragalactic. Their nature remains mysterious, motivating searches for counterparts at other wavelengths. FRB~121102 is so far the only source known to repeatedly emit FRBs and is associated with a host galaxy at redshift $z \simeq 0.193$. We conducted simultaneous observations of FRB~121102 with the Arecibo and MAGIC telescopes during several epochs in 2016--2017. This allowed searches for millisecond-timescale burst emission in very-high-energy (VHE) gamma rays as well as the optical band. While a total of five FRBs were detected during these observations, no VHE emission was detected, neither of a persistent nature nor burst-like associated with the FRBs. The average integral flux upper limits above 100~GeV at 95\% confidence level are $6.6 \times 10^{-12}~\mathrm{photons\ cm^{-2}\ s^{-1}}$ (corresponding to luminosity $L_{\rm VHE} \lesssim 10^{45}~\mathrm{erg\ s^{-1}}$) over the entire observation period, and $1.2 \times 10^{-7}~  \mathrm{photons\ cm^{-2}\ s^{-1}}$ ($L_{\rm VHE} \lesssim 10^{49}~\mathrm{erg\ s^{-1}}$) over the total duration of the five FRBs. We constrain the optical U-band flux to be below 8.6~mJy at 5-$\sigma$ level for 1-ms intervals around the FRB arrival times. A bright burst with U-band flux $29~\mathrm{mJy}$ and duration $\sim 12$~ms was detected 4.3~s before the arrival of one FRB. However, the probability of spuriously detecting such a signal within the sampled time space is 1.5\% (2.2$\upsigma$, post-trial), i.e. consistent with the expected background. We discuss the implications of the obtained upper limits for constraining FRB models.
\end{abstract}




\section{Introduction}

Fast radio bursts (FRBs) are astrophysical phenomena that exhibit bright, transient pulses of millisecond duration, typically at GHz frequencies. First discovered by \citet{FRB_discov}, around 30 such events have been found to date\footnote{See the online FRB Catalog for a list of currently known FRBs:\\ \url{http://www.astronomy.swin.edu.au/pulsar/frbcat/}} \citep{Thornton2013, petroff2016}.
The dispersion measures (DM) observed in FRBs imply intervening column densities of free electrons that are significantly larger than those expected from the Galactic interstellar medium, strongly suggesting their extragalactic origin (c.f. \citealt{Ioka2003,Inoue2004}). However, the nature of FRBs still remains uncertain, mainly because the single-dish radio telescopes used to detect most of these FRBs have localization capabilities that are insufficient for unambiguous identification with counterparts at other wavelengths. A wide variety of theoretical models has been proposed to explain FRBs \citep[see e.g.][for reviews]{katz2018,Rane2017}.

Among the FRB population, only one, FRB~121102, is currently known to exhibit repeating bursts \citep{spitler2014,FRB_repeater,scholz2016}. Its repetitive nature allowed the localization of the source to sub-arcsecond precision, and the discovery of persistent associated sources in the optical and radio bands \citep{chatterjee2017,marcote2017}. It was found to be located in a low-metallicity star-forming region of a dwarf galaxy with $m_{r} = 25.1 \pm 0.1$ mag at a redshift of $z \simeq 0.193$ \citep{tendulkar2017,bassa2017}. The source is localized within a projected separation of 40~pc from a compact ($\lesssim 0.7$ pc) and persistent radio source \citep{marcote2017}.

These findings prompted searches for counterparts of FRB~121102 at other wavelengths. \citet{hardy2017} conducted simultaneous radio and optical observations. Out of a total of 13 radio bursts detected, no significant optical bursts were found above a flux density of 0.33~mJy at 767~nm, corresponding to a fluence limit of 46~mJy~ms. \citet{scholz2017} performed simultaneous radio and X-ray observations. They detected 12 radio bursts, but no X-ray bursts were found in coincidence or at any other epoch, implying $5\sigma$ fluence upper limits of $3 \times 10^{-11}\ \mathrm{erg\ cm^{-2}}$ and $5 \times 10^{-10}\ \mathrm{erg\ cm^{-2}}$ at 0.5 and 10~keV, respectively. No persistent X-ray emission at the position of FRB~121102 was detected. The authors also analyzed {\em Fermi}-GBM data during the epochs of those 12 radio bursts, placing $5\sigma$ fluence upper limits of $4 \times 10^{-9}\ \mathrm{erg\ cm^{-2}}$ ($5 \times 10^{47}\ \mathrm{erg}$ in time-integrated energy at the distance of FRB~121102) in the 10--100~keV energy range. \citet{zhang2017} analysed the eight-year {\em Fermi}-LAT data to search for persistent $\upgamma$-ray emission. No evidence of emission was found, implying an upper limit of $4 \times 10^{44}\ \mathrm{erg\ s^{-1}}$ on the GeV-band luminosity. \cite{bird2017} constrained the persistent very-high-energy (VHE; $\gtrsim 0.1\ \mathrm{TeV}$) emission with VERITAS, setting differential upper limits of $5.2 \times 10^{-12}$ and $4.0 \times 10^{-11}\ \mathrm{cm^{-2}\ s^{-1}\ TeV^{-1}}$ at their energy thresholds of 0.2 and 0.15 TeV (assuming power-law spectra with indices $-2$ and $-4$, respectively).

Various scenarios have been proposed to explain FRB~121102 and the associated persistent radio source. A widely discussed class of FRB progenitors involve neutron stars that are either rotationally powered \citep[e.g.][]{Connor2016,cordes2016,Lyutikov2016} or magnetically powered \citep[e.g.][]{Popov2013,lyubarsky2014}. Pulsar wind nebulae driven by such neutron stars had been predicted as persistent radio counterparts \citep{Murase2016}. \citet{kashiyama2017} showed that a young (10--100~yr old) neutron star powering a pulsar wind nebula inside a supernova remnant could be responsible for FRB~121102. The location of FRB~121102 inside a low-metallicity star-forming region \citep{tendulkar2017} may point to a magnetar, as such environments are similar to the hosts of hydrogen-poor super-luminous supernovae, whose progenitors could be young magnetars \citep{lunnan2014}. On the other hand, \citet{Waxman2017} suggested a self-consistent scenario for both the bursts and the persistent source, in which the associated nebula is surrounded by low-mass ejecta rather than massive ejecta that is often expected for magnetar progenitors \citep[e.g.][]{tendulkar2016, kashiyama2017}.

In a magnetar scenario, quasi-simultaneous X-ray to MeV $\upgamma$-ray bursts could be produced analogously to those observed in short bursts from Galactic magnetars, with a X-ray-to-radio fluence ratio $\sim 10^{4}$ \citep{lyutikov2002}. VHE gamma-ray flashes correlated with FRBs have also been predicted, arising from the interaction of ultra-relativistic outflows triggered by magnetic dissipation with the ambient nebula, with VHE-to-radio fluence ratios $\sim 10^5 \text{--} 10^6$ \citep{lyubarsky2014, Murase2016}. Such VHE emission is possible in certain generic conditions, where a FRB progenitor like a young neutron star or a young white dwarf is naturally surrounded by a hot nebula. If magnetic bursts occur inside the bubble, pre-existing high-energy particles accelerated around the wind termination radius may be accelerated further by the impulsive energy injection into the nebula \citep{Murase2016}. The consequent VHE emission may be detectable when the external shock is strong enough. Such a scheme has also been explored in the synchrotron maser model for FRB emission~\citep{lyubarsky2014}. On the other hand, if the FRBs were caused by coherent curvature radiation, most of the emission may be concentrated in the radio domain, without obvious counterparts at other wavelengths \citep{ghisellini2017}. If FRBs are produced via forced reconnection of magnetic fields near the surface of magnetars, \citet{kumar2017} predicts ms bursts up to optical wavelengths, although independently of FRBs and with a lower burst rate.

An alternative scenario invokes a relativistic jet ejected by a massive black hole (BH) for the origin of both the FRBs and the persistent radio source \citep{vieyro2017}. The luminosity and compactness of the latter may be consistent with a BH with mass $10^4 \text{--} 10^6~\mathrm{M_\odot}$ \citep{marcote2017}. \citet{vieyro2017} suggest that detectable high-energy emission associated with FRBs may occur on timescales of seconds to minutes under certain conditions.

Magnetars can also coexist with a massive BH in the central regions of galaxies \citep[e.g.][]{pen2015,cordes2016}, in which case the former may be responsible for the FRBs and the latter for the persistent radio source. Such systems could be analogous to the magnetars known to exist in the Galactic Center, but with more extreme conditions \citep{pen2015}, possibly even interacting with each other \citep{zhang2018}. The recent discovery of extremely large and variable Faraday rotation of linearly polarized radiation of the bursts from FRB 121102 may be consistent with such environments \citep{Michilli2018}.

To summarize the current knowledge, the progenitors and mechanisms producing FRBs are not well understood, and a variety of predictions have been made for associated counterparts across the electromagnetic spectrum. New and deeper constraints at other wavelengths are necessary to clarify their origin. In this paper, we present optical and VHE observations of FRB~121102 simultaneous with radio observations. The detection of radio bursts during these observations allows us to constrain optical and VHE counterparts correlated in time. 

\S~\ref{sec:obs} describes our simultaneous radio, optical and VHE observations and the data analysis methods. \S~\ref{sec:results} presents the results of the observations. \S~\ref{sec:discussion} discusses the  constraints on the multiwavelength emission of FRB~121102 and the implications. \S~\ref{sec:conclusions} concludes this work.

\section{Instruments, observations and analysis} \label{sec:obs}

In September 2016, we started a campaign of simultaneous observations with the MAGIC (Major Atmospheric Gamma Imaging Cherenkov) telescopes and the Arecibo radio telescope. Observations in VHE gamma rays and in the optical band were carried out with MAGIC, making use of the central pixel installed on the MAGIC II camera \citep{MAGIC_cpix}. We describe below the radio, optical and VHE observations and the data analysis methods.

\subsection{Arecibo radio observations}
\label{sec:arecibo}

Radio observations of FRB~121102 were conducted with the 305-m William E. Gordon Telescope at the Arecibo Observatory at a central frequency of 1.38~GHz. We made use of the Puerto-Rican Ultimate Pulsar Processing Instrument (PUPPI) together with the single-pixel L-band wide receiver, which provide a total bandwidth of 800~MHz and a usable bandwidth of $\sim 600\ \mathrm{MHz}$ (due to radio frequency interference removal).
The data were coherently de-dispersed to $\text{DM} = 557\ \mathrm{pc\ cm^{-3}}$ \citep{spitler2014} to remove the dispersive smearing of the burst widths with a time resolution of $10.24\ \mathrm{\upmu s}$.

A total of five radio bursts from FRB~121102 were detected with a high significance during the simultaneous MAGIC and Arecibo observations using analog methods as in \cite{spitler2014}. We list these bursts and the MAGIC observing conditions at their times of arrival (TOAs) in Table \ref{tab:frb_toas}. We note that the listed Arecibo TOAs refer to the topocentric times on site at the top of the observed band (1.73~GHz). TOAs at the MAGIC site have been corrected to infinite frequency considering the de-dispersion of the signal (776.4~ms) and also corrected for the different expected topocentric times (correction smaller than 10~ms, varying between the different FRBs).

\begin{table*}
\caption{FRBs detected by Arecibo during the campaign, together with the observing conditions for MAGIC at the corresponding epochs. The reported aerosol transmission refers to the atmospheric optical depth relative to a standard dark night. The radio peak brightnesses of the Arecibo bursts have been estimated via the radiometer equation with an uncertainty of $\sim 20\%$ \citep[see][]{scholz2016}.}
\begin{center}
\begin{tabular}{c c c c c|c c c}
\hline
 MJD (Arecibo) & DM & Duration & Peak brightness & Significance (Arecibo) & MJD (MAGIC site) & Aerosol transmission & Zd    \\
 {[}days]   &  [pc cm$^{-3}$]  &  [ms]  & [Jy]  & [$\sigma$]   & [days]           &              & [deg] \\ \hline
 57799.98317566 & 562 & 5.73 & 1.4 & 32.17 & 57799.98316670 & 0.96 & 33 \\ 
 57806.96425078 & 562 & 2.46 & 1.6 & 38.59 & 57806.96424183 & 0.96 & 33 \\ 
 57806.98472905 & 561 & 3.69 & 1.5 & 35.21 & 57806.98472011 & 0.96 & 40 \\ 
 57808.00278585 & 563 & 3.69 & 0.79 & 18.73 & 57808.00277693 & 0.96 & 46 \\ 
 57814.94698520 & 560 & 1.15 & 0.47 & 11.13 & 57814.94697625 & 0.96 & 35 \\
\end{tabular}
\end{center}
\label{tab:frb_toas}
\end{table*}

\subsection{MAGIC observations and data analysis}
\label{sec:observations}

The MAGIC telescope system consists of two 17-m imaging atmospheric Cherenkov telescopes, located at the Roque de los Muchachos Observatory on the island of La Palma, Canary Islands.

\subsubsection{MAGIC stereoscopic observations}
\label{sec:magic_stereo}

Stereoscopic observations with MAGIC provide an integral sensitivity of 0.66 $\pm$ 0.03 \% of the Crab Nebula flux above 220~GeV in 50~h of observation, and allow the measurement of photons in the energy range from 50~GeV to above 50~TeV \citep{MAGIC_performance}.

Observations of FRB~121102, taken up to $60^\circ$ in zenith angle, were carried out in ON mode (i.e. with the source always located at the centre of the field of view) to allow simultaneous data taking in the optical range with the central pixel (see \S \ref{sec:cpix}). The source was observed during 14 nights (between September 2016 and September 2017), with a total of 22~h of data surviving quality cuts, of which 8.9~h were simultaneous with Arecibo. MAGIC observing conditions during the five Arecibo TOAs were excellent, with atmospheric transmission and zenith angles shown in Table \ref{tab:frb_toas}.

The VHE data analysis presented here was carried out using standard MAGIC analysis software \citep{MARS}. Integral and differential flux upper limits were computed as in \cite{Rolke} assuming a 30\% systematic uncertainty on the efficiency. Given that ON-mode observations do not allow to use the standard background evaluation methods (simultaneous background using reflected regions, see \citealt{MAGIC_performance}), the background was extracted from OFF data samples collected under similar conditions (mainly zenith angle and night-sky background level).

For the search of millisecond-timescale VHE emission described in \S \ref{sec:ms_vhe}, integral flux upper limits were calculated assuming that the expected number of photons within a 10-ms time window follows a Poisson distribution with no expected background (see for instance Table 39.3 in \cite{pdg_2016}). A toy Monte Carlo simulation was performed to estimate the flux upper limit under the assumption of a 30\% Gaussian systematic uncertainty on the mean.

\subsubsection{MAGIC central pixel}
\label{sec:cpix}

The MAGIC telescopes are able to operate simultaneously as both VHE and optical telescopes, with excellent sensitivity in the two regimes. Optical observations are performed using only one pixel within one of the MAGIC cameras, namely the central pixel, which covers a 0.1 deg field of view. It consists of a fully modified photosensor-to-readout chain at the centre of the MAGIC-II telescope camera, increasing the bandwidth of the central pixel DC branch from 8 Hz to over 3 kHz \citep{cpix_icrc}. After the upgrade carried out in 2011-2012 \citep{upgrade1}, MAGIC is able to detect the optical pulsations of the Crab Pulsar with observation times shorter than 10~s \citep{cpix_icrc}. By studying the dispersion within the off-source data and making use of the well known flux and phaseogram of the Crab Pulsar, the MAGIC central pixel is able to detect isolated 1-ms optical flashes as faint as $\sim 8$~mJy (13.4 mag) with maximum sensitivity at 350~nm \citep{cpix_icrc, MAGIC_II}.
The central pixel data exhibits some low frequency noise, mainly caused by surrounding camera components. In order to improve sensitivity to 1\text{--}10~ms non-periodic optical pulses, averaging filters of variable integration length (1\text{--}10~ms) were applied to denoise the central pixel data with a standard sampling rate of 10 kHz. At each point, the uncertainty is taken as the standard deviation of each 1\text{--}10~ms window.

Various external light sources such as meteors, car flashes, satellites, and space debris are able to produce fast optical pulses, constituting backgrounds in searching for optical counterparts to FRBs. Several methods to identify these events have been developed and implemented for the data analysis presented here, mainly involving the use of average pixel DC current reports that are stored every second during observations. Identifying variations in the average camera currents on timescales of seconds to minutes efficiently removes slowly varying optical signals (e.g. car flashes or satellites). However, after the use of these filters, an irreducible background still remains, mainly produced by faint meteors passing through the field of view of the central pixel, producing signals lasting about 5 to 20 ms. 

The frequency of these background events was studied to allow the calculation of the significance of a hypothetical non-simultaneous optical pulse that may precede or follow the radio bursts. Within each off-source data run, after applying a 1-ms averaging filter and the selection criteria described above, any signal exceeding 5 times the standard deviation of the average voltage was classified as a possible optical background pulse. Even if the background rate is low (frequency between 10$^{-2} \text{--} 10^{-5}$ Hz, decreasing with brightness), it hinders the possibility of associating with high confidence an optical burst that is not precisely simultaneous with a given FRB.

The central pixel sensitivity and timing precision was tested every night of FRB~121102 observations by dedicating five minutes to observe the Crab Pulsar. To convert the central pixel output voltage to the corresponding optical absolute magnitude, an empirical expression was derived by fitting the measured phaseogram of the Crab Pulsar to its well known flux profile, as done in \cite{cpix_icrc}. The magnitude in the U band measured by the central pixel can be expressed as

\begin{equation}
	\label{eq:cpix_conversion}
	m_{\rm U} = m_{\rm U, CP} - 2.5 \log_{10}(2.15 \times 10^3 V - 2.28 \times 10^2)
\end{equation}
where $m_{\rm U, CP}$ is the average optical magnitude of the Crab Pulsar in the U band ($\sim$ 16.9 mag) and $V$ is the output voltage in volts.

\section{Results} \label{sec:results}

Given the large variety of predictions available for counterparts across the electromagnetic spectrum, several kinds of searches have been performed with the data sample described in \S~\ref{sec:observations}. We searched for millisecond-timescale burst emission associated with the FRBs detected by Arecibo, both in the optical and VHE range, as well as persistent VHE gamma-ray emission as in \cite{bird2017}.

\subsection{Persistent VHE emission}

\begin{figure}
\begin{center}
\includegraphics[width=0.47\textwidth]{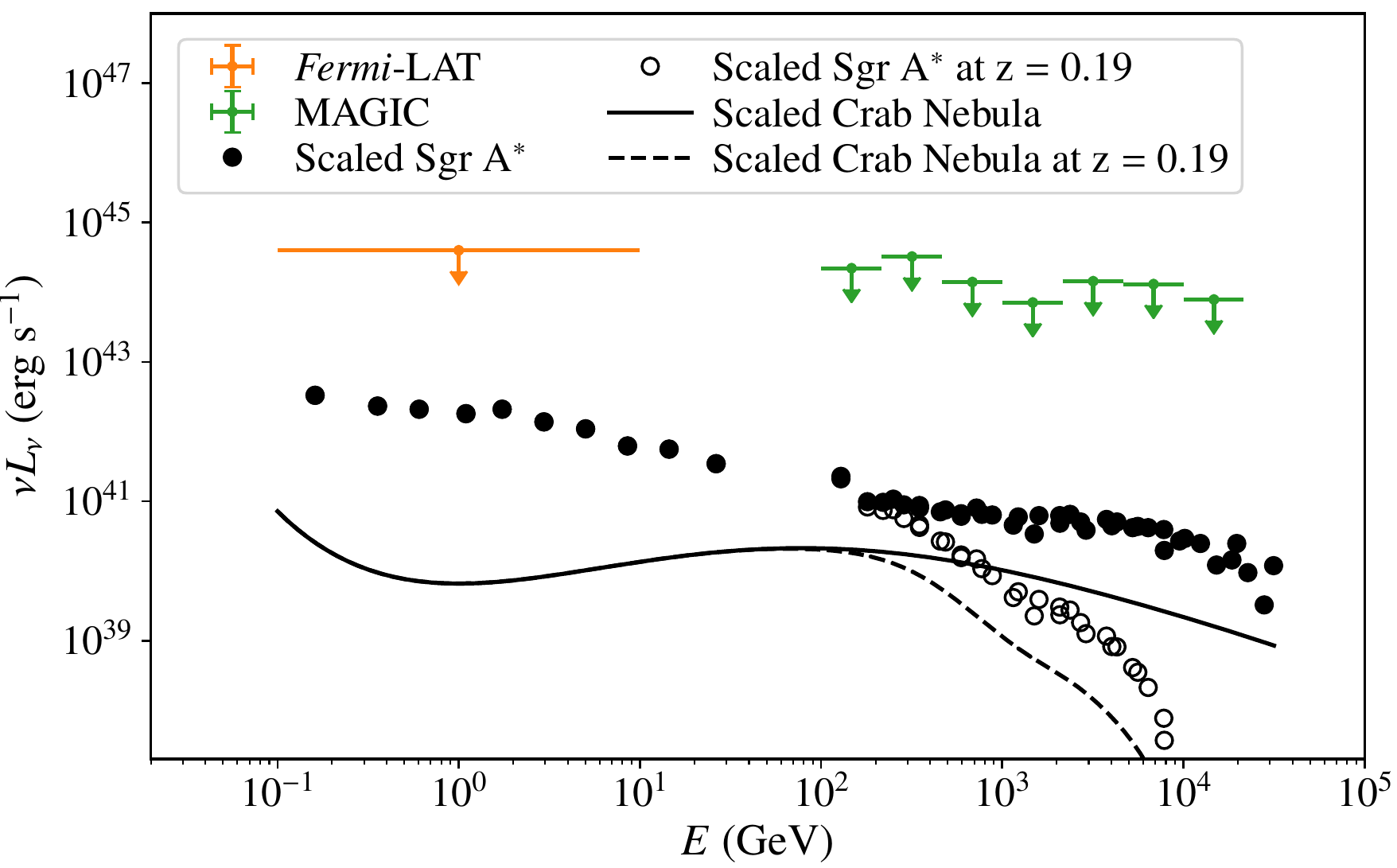}
\caption{Upper limits in luminosity for the persistent gamma-ray emission of FRB~121102 from MAGIC (95\% confidence level, assuming an intrinsic power-law spectrum with $\Gamma = 2$ and 30\% overall systematic uncertainty). Limits from {\em Fermi}-LAT \citep{zhang2017} are also shown. The black curve and filled circles represent, respectively, the SEDs of the Crab Nebula \citep{meyer2010} and Sgr~A$^\star$ \citep{lat_bright_sources, HESS_SgrA_2009}, scaled by factors of $4 \times 10^5$ and $2 \times 10^6$ to match the observed radio luminosity of the persistent counterpart of FRB 121102 (see \S \ref{sec:vhe_persistent}). The dashed black curve and empty circles show the effect of gamma-ray attenuation from \mbox{z = 0.19} due to the EBL, following \citet{dominguez11a}.}
\label{fig:vhe_sed}
\end{center}
\end{figure}

No persistent VHE gamma-ray emission was detected from FRB~121102. Assuming a power-law spectrum with photon index $\Gamma$, integral flux upper limits (ULs) were calculated above 100, 400 and 1000 GeV at 95\% confidence level. These results are shown in Table~\ref{tab:ULs_persistent}. For the specific case of $\Gamma = 2$ assumed for each energy bin, differential flux upper limits are shown in Fig.~\ref{fig:vhe_sed}.

\begin{table}
\caption{Upper limits on the persistent VHE emission of FRB~121102 in terms of integral flux above $E_{0}$, assuming power-law spectra with $\Gamma = 2$ and $\Gamma = 4$.}
\begin{center}
\begin{tabular}{ c c c }
\hline
$E_{0}$  & Int. flux UL ($\Gamma = 2$) & Int. flux UL ($\Gamma = 4$) \\ 
   {[}GeV]      & [$10^{-12}\ \mathrm{cm^{-2}\ s^{-1}}$] & [$10^{-12}\ \mathrm{cm^{-2}\ s^{-1}}$] \\ \hline
100 & 6.6 & 12\\
400 & 1.7 & 1.9\\
1000 & 0.37 & 0.33 \\
\end{tabular}
\end{center}
\label{tab:ULs_persistent}
\end{table}

\subsection{Millisecond-timescale VHE emission}
\label{sec:ms_vhe}

Fixing a time window of 10~ms centred around the radio burst TOAs and using custom analysis cuts (on \textit{size}, \textit{Hadronness} and \textit{$\theta^2$}; see \citealt{MAGIC_performance}) that are optimized to maximize sensitivity for a 10-ms signal, no gamma-like events are found within any of these windows above 100~GeV. Since the background rate during such time intervals is negligible with <10$^{-2}$, the resulting UL (95\% CL, adding 30\% systematic uncertainty to the Poisson mean) for each FRB corresponds to 3.56 events (see \S \ref{sec:magic_stereo}). The corresponding integral flux upper limits for individual FRBs in different energy ranges are shown in Table \ref{tab:ULs_ms}, assuming a power-law spectrum with two different indices, $\Gamma = 2$ and $\Gamma = 4$. A combined integral flux upper limit is also derived by stacking the data around the 5 FRB TOAs, over a duration of $5 \times 10$~ms.

\begin{table}
\caption{Upper limits on VHE burst emission of FRB 121102, in terms of integral flux above $E_{0}$ over 10~ms intervals around the TOAs of each FRB, assuming power-law spectra with $\Gamma = 2$ and $\Gamma = 4$. Limits on the average flux over 50 ms are also shown, derived by combining the data for the 5 FRBs. These limits are also valid for shorter integration time windows.}
\begin{center}
\begin{tabular}{ c c c c }
\hline
FRB MJD & $E_{0}$  & Int. flux UL ($\Gamma = 2$) & Int. flux UL ($\Gamma = 4$) \\
{[}days]                     &    [GeV]      & [$10^{-7}\ \mathrm{ cm^{-2}\ s^{-1}}$] & [$10^{-7}\ \mathrm{ cm^{-2}\ s^{-1}}$] \\ \hline
57799.98 	& 100 	& 5.7 	& 9.3 \\
         	& 400 	& 2.9 	& 3.1 \\
         	& 1000 	& 2.5 	& 2.2 \\
57806.96 	& 100 	& 5.7 	& 9.3 \\
         	& 400 	& 2.9 	& 3.1 \\
         	& 1000 	& 2.5 	& 2.2 \\
57806.98 	& 100 	& 5.6 	& 10 \\
         	& 400 	& 2.6 	& 2.9 \\
         	& 1000 	& 2.1 	& 1.7 \\
57808.00 	& 100 	& 5.6 	& 14 \\
         	& 400 	& 2.2 	& 2.6 \\
         	& 1000 	& 1.7 	& 1.4 \\
57814.95 	& 100 	& 5.5 	& 8.5 \\
         	& 400 	& 2.8 	& 2.8 \\
         	& 1000 	& 2.4 	& 2.2 \\ \hline
Combined 	& 100 	& 1.2 	& 2.3 \\
        	& 400 	& 0.52 	& 0.59 \\
        	& 1000 	& 0.41 	& 0.36 \\

\end{tabular}
\end{center}
\label{tab:ULs_ms}
\end{table}

Given that the properties of FRBs are unknown, offsets in the arrival times of the burst emission at radio and higher frequency are possible. A blind search for non-simultaneous VHE bursts was also performed. In this case, due to the large number of trials, the sensitivity worsens significantly. A (non-overlapping) sliding 10-ms window sampling the arrival time of all events surviving analysis cuts was used through the whole dataset ($1.2 \times 10^5$ trials per 20-min run).

No hint of VHE bursts was found, for any offset up to an hour with respect to the Arecibo TOAs. The minimum flux of VHE photons detectable by this blind search was calculated from the joint probability density function of all gamma-like events observed, corrected by the total number of trials performed. We conclude that a single 10-ms burst with a flux of $8.2 \times 10^{-5}~\mathrm{cm^{-2}\ s^{-1}}$ above 100~GeV (equivalent to an isotropic luminosity of $L_{\rm VHE} \sim 10^{52} \mathrm{erg\ s^{-1}}$)  would have been firmly detected (S > 5 $\sigma$), assuming power-law spectra with $\Gamma = 2$.

\subsection{Millisecond-timescale optical emission}

\begin{figure}
\begin{center}
\includegraphics[width=0.53\textwidth]{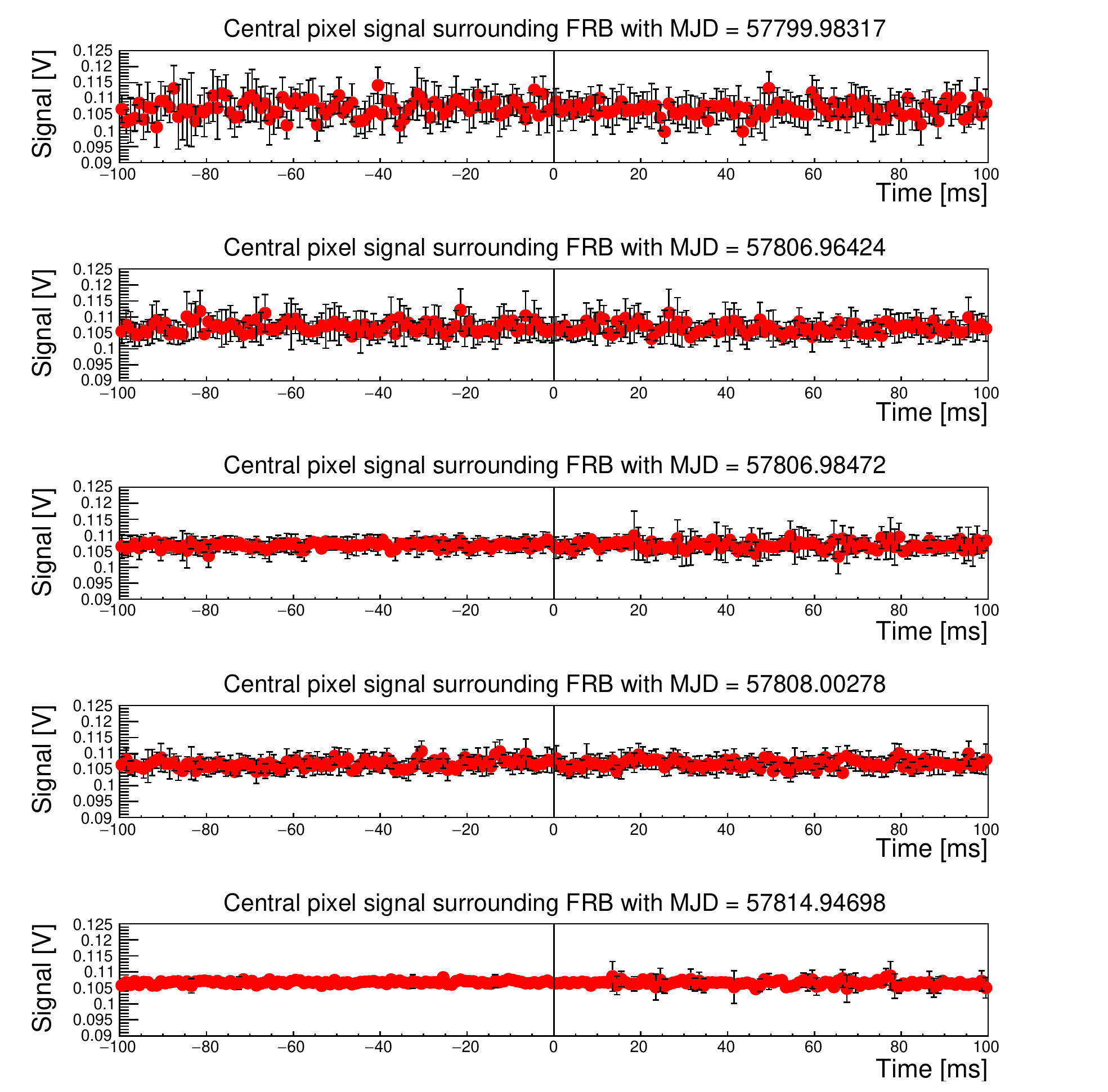}
\caption{Optical light curves covering 200~ms around the TOAs of the 5 radio bursts from FRB~121102 detected by the Arecibo telescope simultaneous with MAGIC data, for an integration window of 1~ms. The vertical axis is proportional to the U-band flux. No significant excess is observed simultaneously with any of the 5 bursts. The noise level varies with the sky brightness.}
\label{fig:cpix_lightcurves}
\end{center}
\end{figure}

As described in \S~\ref{sec:cpix}, observations were carried out in the optical U-band using the MAGIC central pixel. As shown in Fig. \ref{fig:cpix_lightcurves}, no significant excess is detected simultaneously with any of the 5 FRB events. As discussed in \cite{cpix_icrc}, the sensitivity of the central pixel varies depending on the assumed duration of the signal. For integration times of 0.1, 1, 5 and 10~ms, the 5-$\sigma$ sensitivity is 20, 8.6, 4.2 and 3.2~mJy, respectively. The sensitivity averaged over the 5 FRB events can also be derived by stacking the data around their TOAs, giving 12, 4.1, 2.3 and 1.7~mJy for the same integration times as above.

As introduced in \S~\ref{sec:cpix}, the irreducible background of optical pulses within our OFF data sample hinders the search for optical bursts with arrival times offset from the radio TOAs. Nevertheless, due to the relatively low frequency of expected background events, searches for such optical pulses are worthwhile as long as the search window is sufficiently small. Thus, an unbiased search for 10-ms optical pulses around the radio TOAs was conducted, sequentially increasing the total search window around each FRB in equal logarithmic time steps (starting from 10~ms, then 100~ms, 1~s and so on). The number of trials for such a pulse search would correspond to the total number of 10-ms bins within the search window of all the FRB events scrutinized (e. g. if 1 second around each TOA was sampled, the total number of trials would correspond to N = 5 TOAs $\times$ 1 s $\times$ 100 trials/s).

\begin{figure}
\begin{center}
\includegraphics[width=0.51\textwidth]{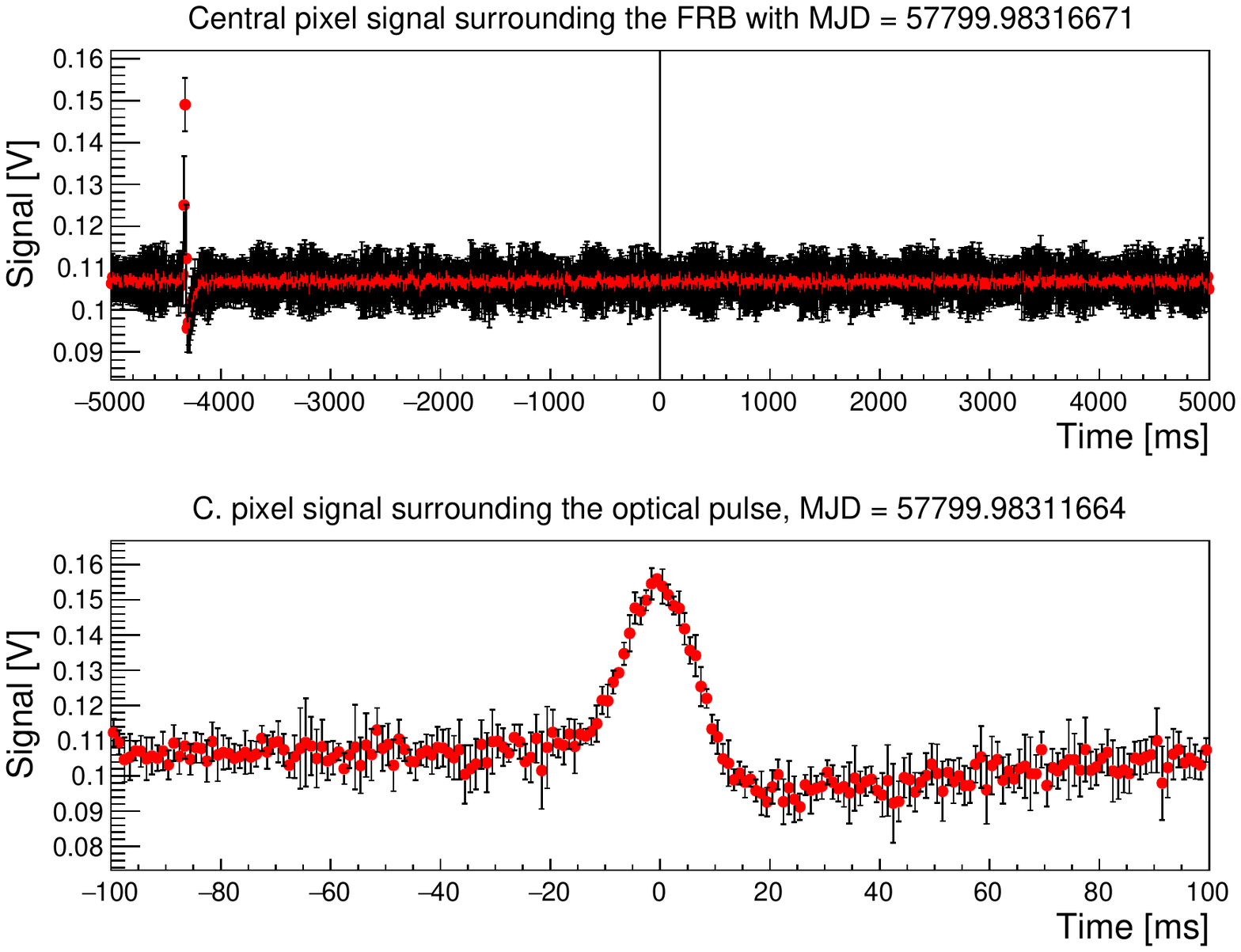}
\caption{\textit{Top}: Optical light curve covering 10~s around the first FRB in our sample, for an integration window of 10~ms. A clear optical pulse is detected 4.3~s before the FRB. \textit{Bottom}: Optical light curve covering 200~ms around the detected optical pulse, for an integration window of 1~ms. The pulse is consistent with a background event. Note that the undershoot after the optical flash is caused by the central pixel readout electronics.}
\label{fig:optical_hint}
\end{center}
\end{figure}

A bright optical pulse with a peak brightness of \mbox{$\sim 29$~mJy} and FWHM of 12 ms was clearly detected 4.3~s before the first FRB in our sample (Fig.~\ref{fig:optical_hint}). No optical pulses are detected near the TOA of any other radio burst. Taking into account the frequency of pulses surviving analysis cuts within our OFF source data sample (a total of 17 within 15.5~h), an optical pulse of this brightness is consistent with the observed background. The resulting chance probability is 1.5\% post-trial. The time profile of this optical pulse is consistent with that of background pulses. For reference, the aforementioned Arecibo burst exhibited a radio peak brightness of 1.4~Jy (see Table~\ref{tab:frb_toas}).

\section{Discussion} \label{sec:discussion}

We have reported on 22 h of VHE gamma-ray and optical observations of FRB~121102 with MAGIC. Simultaneous radio observations were conducted with Arecibo for 8.9 h, revealing a total of 5 radio bursts. We have derived constraints on the VHE and optical burst emission as well as the persistent VHE emission. Below we discuss some physical implications in light of potential scenarios for FRB 121102.

\subsection{Persistent VHE emission}
\label{sec:vhe_persistent}

The origin of the persistent radio source associated with FRB~121102 remains unclear. One possibility is a pulsar wind nebula driven by a young rotation-powered neutron star or magnetar (\citealt{kashiyama2017}; see however \citealt{Waxman2017}), which may also emit persistent VHE gamma rays via mechanisms analogous to known pulsar wind nebulae \citep{Murase2016}. Alternatively, it could be related to a BH with mass $\sim 10^4 \text{--} 10^6\ \mathrm{M_{\odot}}$ \citep{marcote2017}, which may have associated persistent VHE emission similar to that observed from Sgr~A$^\star$ in the Galactic Center.

Figure~\ref{fig:vhe_sed} shows the upper limits for persistent HE to VHE gamma rays from FRB~121102. These are compared with SEDs of the Crab Nebula and Sgr~A$^\star$, that have been scaled respectively by factors of $4\times 10^5$ and $2 \times 10^6$ to match the observed radio luminosity of the persistent radio source associated with FRB 121102. Also shown in Figure~\ref{fig:vhe_sed} is the effect of attenuation of the gamma rays by $\gamma\gamma$ pair production interactions with the extragalactic background light (EBL), which is significant above $\sim$ 400 GeV at the redshift of FRB 121102 \citep[e.g.][]{dominguez11a}.

The current upper limits for persistent gamma-ray emission lie $\sim 2$--$4$ orders of magnitude above such simple expectations based on luminosity scaling, and cannot provide significant constraints on the nature of FRB~121102.

\subsection{VHE burst emission associated with FRBs}

As shown in Table~\ref{tab:ULs_ms}, the obtained upper limits on the burst-like VHE photon flux simultaneous with the FRBs are in the range $\sim (3\text{--}10) \times 10^{-7}~\mathrm{cm^{-2}\ s^{-1}}$ above $100\text{--}400$ GeV over durations of 10~ms. With the luminosity distance $d_{\rm L} = 972 \ \rm Mpc$ for FRB 121102, this implies limits on the VHE luminosity per burst of $L_{\rm VHE} \lesssim (3\text{--}14) \times 10^{49}~\mathrm{erg\ s^{-1}}$. Also, from the limits on the flux averaged over the FRBs during 50~ms, the upper limit on the VHE radiation energy per burst can be roughly estimated as ${\mathcal E}_{\rm VHE} \lesssim (3\text{--}9) \times 10^{47}$ erg. Limits above 1 TeV would be a factor $\sim$ 10 less constraining due to the effect of attenuation by the EBL.

Such upper limits can provide a valuable test of the magnetar scenario by constraining the burst energy carried by re-accelerated electron-positron pairs \citep{lyubarsky2014,Murase2016}. If each FRB results from release of the magnetic free energy ${\mathcal E}_B$ trapped in the neutron star magnetosphere, a highly relativistic magnetized outflow will be launched. When such an outflow interacts with the slower nebula and its energy is dissipated, pre-existing non-thermal electrons and positrons are accelerated and emit synchrotron and inverse-Compton emission, which may be observable as a broadband flare from the radio to VHE bands. The energy dissipation timescale is highly uncertain and may range from a few ms to much longer, depending on e.g. the Lorentz factor of the outflow~\citep{Murase2016}. In the fast cooling regime where electrons and positrons cool within the dynamical time, the radiated energy at VHE can be as large as ${\mathcal E}_{\rm VHE} \sim 10^{47}~\mathrm{erg} \ ({\cal C}/10)^{-1} ({\mathcal E}_B/10^{48}~\mathrm{erg})$, where ${\cal C} = O(10)$ is a factor that accounts for bolometric correction. Our results imply that the released magnetic energy may be constrained to be ${\mathcal E}_{\rm B} \lesssim 10^{48}$ erg in the fast-cooling limit. 

The current constraint on the energetics is not very stringent but can be significantly improved in the future. As the number $N_{\rm FRB}$ of observed FRBs increases, the constraint would become tighter by a factor of about $N_{\rm FRB}$ in the background-free limit, or $N_{\rm FRB}^{1/2}$ if the background is non-negligible. For example, if the MAGIC telescopes could obtain similar limits for $N_{\rm FRB} = 500$ repeating bursts from FRB~121102, the upper limit on the released energy will reach ${\mathcal E}_{\rm B} \lesssim 10^{46}$ erg, comparable to that of known magnetar hyper-flares \citep{Popov2013,lyubarsky2014}. Such observations by current Cherenkov telescopes may be feasible over a period of $\sim 5\text{--}10$ years, given the frequency of bursts observed from FRB~121102 during some periods that can be as large as 18 bursts in 30 min \cite{gajjar_2018}. If other sources of repeating FRBs are found that are more nearby, the constraints could be tightened by a factor $\sim d_{\rm L}^{2}$ at $\sim$ 100 GeV, and by a larger factor at higher energies by virtue of the reduced EBL attenuation. Further drastic improvements can be expected with future radio and VHE observatories, such as the Square Kilometer Array\footnote{\url{https://www.skatelescope.org/}} and the Cherenkov Telescope Array\footnote{\url{https://www.cta-observatory.org/}}. Thus, VHE observations simultaneous with radio bursts provide a potentially powerful test of the magnetar model for FRB progenitors.

\subsection{Optical burst emission associated with FRBs}

No optical bursts on ms timescales coincident with radio bursts have been detected so far. We constrain the optical flux to be below 4.1 and 1.7~mJy at 5-$\sigma$ confidence level for 1 and 10-ms time windows around the radio TOAs, respectively. For optical bursts unassociated with radio bursts, limits of 8.6 and 3.2~mJy are obtained for 1 and 10-ms duration, respectively. We note that a significant fraction of individual bursts from FRB~121102 are observed to emit within a relatively narrow frequency range in the radio domain \citep{law2017}. Thus, correlations between radio and optical bursts may possibly be weak. Compared to previously reported upper limits on the fluence at 767~nm of 46~mJy~ms \citep{hardy2017}, our upper limits provide the most stringent constraints on the putative optical burst emission of FRB~121102 to date. This can be contrasted with the Crab Pulsar, for which coincident radio and optical pulses have been reported \citep{Shearer2003} with a radio-to-optical flux density slope (energy index) of $\alpha \sim - 0.2$ \citep[see e.g.][]{Lyne2005}. For FRB~121102, the optical upper-limits presented here provide a strong constraint on this slope of $\alpha \lesssim - 0.32$.

As shown in Fig. \ref{fig:optical_hint}, we have detected an optical pulse with peak brightness of 28.9~mJy that arrived $\sim 4.3$~s before a 1.4-Jy radio burst, for which the TOA has been corrected for dispersion at infinite frequency and topocentric time. Optical pulses preceding radio pulses have been previously observed in giant radio pulses from the Crab Pulsar \citep{Shearer2003,Strader2013}. If we assume that the optical pulse and radio burst of FRB~121102 are physically connected, it would point to a slope between the optical and radio flux of $\alpha \sim -0.3$, similar to but slightly steeper than that observed in the Crab pulses ($\alpha \sim - 0.2$). However, the detected optical pulse is compatible with the time profile and brightness of known background signals such as meteors and we cannot ascertain its origin. Further observations by MAGIC and other high time-resolution optical telescopes will provide stronger tests of potential optical burst emission from FRB~121102.

\section{Conclusions} \label{sec:conclusions}

We have conducted simultaneous radio, optical and VHE observations of FRB~121102 with the Arecibo and MAGIC telescopes, in order to search for burst emission on millisecond timescales at these wavelengths. For the first time, we constrain the VHE and optical burst emission simultaneous with FRBs, 5 of which were detected during our campaign. We obtain limits of $0.5 \times 10^{-7}~\mathrm{cm^{-2}\ s^{-1}}$ for the average flux above 100 GeV during the duration of the 5 FRBs, with interesting future implications for constraining some FRB models involving magnetars. We also set limits of 8.6~mJy for the U-band flux during 1-ms intervals around the FRB arrival times, the strongest such constraints to date. We also obtain limits on the persistent VHE emission comparable to that already reported by \cite{bird2017}, which are still $\sim 2$--$4$ orders of magnitude above simple expectations based on scaling the SEDs of well known sources such as the Crab Nebula and Sgr~A$^\star$.

The optical pulse observed 4.3~s before the first FRB detected by Arecibo during our campaign cannot be unambiguously associated with FRB 121102 (being at 2.2 $\sigma$ confidence level). This is consistent with the fact that no other bright optical pulse has been found within a few seconds of the other four FRBs. However, it is worth noting that the spectra and time profiles of FRBs are known to be extremely variable from burst to burst. It is possible that only a limited number of these bursts are sufficiently bright to be detectable at higher frequencies, encouraging more searches for optical pulses.

\section*{Acknowledgements}
%
%
We would like to thank the Instituto de Astrof\'{\i}sica de Canarias for the excellent working conditions at the Observatorio del Roque de los Muchachos in La Palma. The financial support of the German BMBF and MPG, the Italian INFN and INAF, the Swiss National Fund SNF, the ERDF under the Spanish MINECO (FPA2015-69818-P, FPA2012-36668, FPA2015-68378-P, FPA2015-69210-C6-2-R, FPA2015-69210-C6-4-R, FPA2015-69210-C6-6-R, AYA2015-71042-P, AYA2016-76012-C3-1-P, ESP2015-71662-C2-2-P, CSD2009-00064), and the Japanese JSPS and MEXT is gratefully acknowledged. This work was also supported by the Spanish Centro de Excelencia ``Severo Ochoa'' SEV-2012-0234 and SEV-2015-0548, and Unidad de Excelencia ``Mar\'{\i}a de Maeztu'' MDM-2014-0369, by the Croatian Science Foundation (HrZZ) Project IP-2016-06-9782 and the University of Rijeka Project 13.12.1.3.02, by the DFG Collaborative Research Centers SFB823/C4 and SFB876/C3, the Polish National Research Centre grant UMO-2016/22/M/ST9/00382 and by the Brazilian MCTIC, CNPq and FAPERJ. SI thanks support from JSPS KAKENHI Grant Number JP17K05460. The work of K.M. is partially supported by Alfred P. Sloan Foundation and NSF grant No. PHY-1620777.




\bibliographystyle{mnras}
\bibliography{references} 

\begin{thebibliography}{}
\makeatletter
\relax
\def\mn@urlcharsother{\let\do\@makeother \do\$\do\&\do\#\do\^\do\_\do\%\do\~}
\def\mn@doi{\begingroup\mn@urlcharsother \@ifnextchar [ {\mn@doi@}
  {\mn@doi@[]}}
\def\mn@doi@[#1]#2{\def\@tempa{#1}\ifx\@tempa\@empty \href
  {http://dx.doi.org/#2} {doi:#2}\else \href {http://dx.doi.org/#2} {#1}\fi
  \endgroup}
\def\mn@eprint#1#2{\mn@eprint@#1:#2::\@nil}
\def\mn@eprint@arXiv#1{\href {http://arxiv.org/abs/#1} {{\tt arXiv:#1}}}
\def\mn@eprint@dblp#1{\href {http://dblp.uni-trier.de/rec/bibtex/#1.xml}
  {dblp:#1}}
\def\mn@eprint@#1:#2:#3:#4\@nil{\def\@tempa {#1}\def\@tempb {#2}\def\@tempc
  {#3}\ifx \@tempc \@empty \let \@tempc \@tempb \let \@tempb \@tempa \fi \ifx
  \@tempb \@empty \def\@tempb {arXiv}\fi \@ifundefined
  {mn@eprint@\@tempb}{\@tempb:\@tempc}{\expandafter \expandafter \csname
  mn@eprint@\@tempb\endcsname \expandafter{\@tempc}}}

\bibitem[\protect\citeauthoryear{{Abdo} et~al.,}{{Abdo}
  et~al.}{2009}]{lat_bright_sources}
{Abdo} A.~A.,  et~al., 2009, \mn@doi [\apjs] {10.1088/0067-0049/183/1/46},
  \href {http://adsabs.harvard.edu/abs/2009ApJS..183...46A} {183, 46}

\bibitem[\protect\citeauthoryear{{Aharonian} et~al.}{{Aharonian}
  et~al.}{2009}]{HESS_SgrA_2009}
{Aharonian} F.,  et~al., 2009, \mn@doi [\aap] {10.1051/0004-6361/200811569},
  \href {http://cdsads.u-strasbg.fr/abs/2009A%26A...503..817A} {503, 817}

\bibitem[\protect\citeauthoryear{{Aleksi{\'c}} et~al.}{{Aleksi{\'c}}
  et~al.}{2016a}]{upgrade1}
{Aleksi{\'c}} J.,  et~al., 2016a, \mn@doi [Astroparticle Physics]
  {10.1016/j.astropartphys.2015.04.004}, \href
  {http://adsabs.harvard.edu/abs/2016APh....72...61A} {72, 61}

\bibitem[\protect\citeauthoryear{{Aleksi{\'c}} et~al.}{{Aleksi{\'c}}
  et~al.}{2016b}]{MAGIC_performance}
{Aleksi{\'c}} J.,  et~al., 2016b, \mn@doi [Astroparticle Physics]
  {10.1016/j.astropartphys.2015.02.005}, \href
  {http://adsabs.harvard.edu/abs/2016APh....72...76A} {72, 76}

\bibitem[\protect\citeauthoryear{{Bassa} et~al.,}{{Bassa}
  et~al.}{2017}]{bassa2017}
{Bassa} C.~G.,  et~al., 2017, \mn@doi [\apjl] {10.3847/2041-8213/aa7a0c}, \href
  {http://adsabs.harvard.edu/abs/2017ApJ...843L...8B} {843, L8}

\bibitem[\protect\citeauthoryear{{Bird} et~al.}{{Bird} et~al.}{2017}]{bird2017}
{Bird} R.,  et~al., 2017, preprint, \href
  {http://adsabs.harvard.edu/abs/2017arXiv170804717B} {} (\mn@eprint {arXiv}
  {1708.04717})

\bibitem[\protect\citeauthoryear{{Borla Tridon} et~al.,}{{Borla Tridon}
  et~al.}{2009}]{MAGIC_II}
{Borla Tridon} D.,  et~al., 2009, preprint, \href
  {http://adsabs.harvard.edu/abs/2009arXiv0906.5448B} {} (\mn@eprint {arXiv}
  {0906.5448})

\bibitem[\protect\citeauthoryear{{Chatterjee} et~al.,}{{Chatterjee}
  et~al.}{2017}]{chatterjee2017}
{Chatterjee} S.,  et~al., 2017, \mn@doi [\nat] {10.1038/nature20797}, \href
  {http://adsabs.harvard.edu/abs/2017Natur.541...58C} {541, 58}

\bibitem[\protect\citeauthoryear{{Connor}, {Sievers}  \& {Pen}}{{Connor}
  et~al.}{2016}]{Connor2016}
{Connor} L.,  {Sievers} J.,   {Pen} U.-L.,  2016, \mn@doi [\mnras]
  {10.1093/mnrasl/slv124}, \href
  {http://adsabs.harvard.edu/abs/2016MNRAS.458L..19C} {458, L19}

\bibitem[\protect\citeauthoryear{{Cordes} \& {Wasserman}}{{Cordes} \&
  {Wasserman}}{2016}]{cordes2016}
{Cordes} J.~M.,  {Wasserman} I.,  2016, \mn@doi [\mnras]
  {10.1093/mnras/stv2948}, \href
  {http://adsabs.harvard.edu/abs/2016MNRAS.457..232C} {457, 232}

\bibitem[\protect\citeauthoryear{{Dom{\'{\i}}nguez} et~al.}{{Dom{\'{\i}}nguez}
  et~al.}{2011}]{dominguez11a}
{Dom{\'{\i}}nguez} A.,  et~al., 2011, \mn@doi [MNRAS]
  {10.1111/j.1365-2966.2010.17631.x}, \href
  {http://adsabs.harvard.edu/abs/2011MNRAS.410.2556D} {410, 2556}

\bibitem[\protect\citeauthoryear{{Gajjar} et~al.,}{{Gajjar}
  et~al.}{2018}]{gajjar_2018}
{Gajjar} V.,  et~al., 2018, preprint, \href
  {http://adsabs.harvard.edu/abs/2018arXiv180404101G} {} (\mn@eprint {arXiv}
  {1804.04101})

\bibitem[\protect\citeauthoryear{{Ghisellini} \& {Locatelli}}{{Ghisellini} \&
  {Locatelli}}{2017}]{ghisellini2017}
{Ghisellini} G.,  {Locatelli} N.,  2017, preprint, \href
  {http://adsabs.harvard.edu/abs/2017arXiv170807507G} {} (\mn@eprint {arXiv}
  {1708.07507})

\bibitem[\protect\citeauthoryear{{Hardy} et~al.,}{{Hardy}
  et~al.}{2017}]{hardy2017}
{Hardy} L.~K.,  et~al., 2017, \mn@doi [\mnras] {10.1093/mnras/stx2153}, \href
  {http://adsabs.harvard.edu/abs/2017MNRAS.472.2800H} {472, 2800}

\bibitem[\protect\citeauthoryear{{Hassan} et~al.,}{{Hassan}
  et~al.}{2017}]{cpix_icrc}
{Hassan} T.,  et~al., 2017, preprint, \href
  {http://adsabs.harvard.edu/abs/2017arXiv170807698H} {} (\mn@eprint {arXiv}
  {1708.07698})

\bibitem[\protect\citeauthoryear{{Inoue}}{{Inoue}}{2004}]{Inoue2004}
{Inoue} S.,  2004, \mn@doi [\mnras] {10.1111/j.1365-2966.2004.07359.x}, \href
  {http://adsabs.harvard.edu/abs/2004MNRAS.348..999I} {348, 999}

\bibitem[\protect\citeauthoryear{{Ioka}}{{Ioka}}{2003}]{Ioka2003}
{Ioka} K.,  2003, \mn@doi [\apjl] {10.1086/380598}, \href
  {http://adsabs.harvard.edu/abs/2003ApJ...598L..79I} {598, L79}

\bibitem[\protect\citeauthoryear{{Kashiyama} \& {Murase}}{{Kashiyama} \&
  {Murase}}{2017}]{kashiyama2017}
{Kashiyama} K.,  {Murase} K.,  2017, \mn@doi [\apjl]
  {10.3847/2041-8213/aa68e1}, \href
  {http://adsabs.harvard.edu/abs/2017ApJ...839L...3K} {839, L3}

\bibitem[\protect\citeauthoryear{{Katz}}{{Katz}}{2018}]{katz2018}
{Katz} J.~I.,  2018, preprint, \href
  {http://adsabs.harvard.edu/abs/2018arXiv180409092K} {} (\mn@eprint {arXiv}
  {1804.09092})

\bibitem[\protect\citeauthoryear{{Kumar}, {Lu}  \& {Bhattacharya}}{{Kumar}
  et~al.}{2017}]{kumar2017}
{Kumar} P.,  {Lu} W.,   {Bhattacharya} M.,  2017, \mn@doi [\mnras]
  {10.1093/mnras/stx665}, \href
  {http://adsabs.harvard.edu/abs/2017MNRAS.468.2726K} {468, 2726}

\bibitem[\protect\citeauthoryear{{Law} et~al.,}{{Law} et~al.}{2017}]{law2017}
{Law} C.~J.,  et~al., 2017, preprint, \href
  {http://adsabs.harvard.edu/abs/2017arXiv170507553L} {} (\mn@eprint {arXiv}
  {1705.07553})

\bibitem[\protect\citeauthoryear{{Lorimer}, {Bailes}, {McLaughlin}, {Narkevic}
  \& {Crawford}}{{Lorimer} et~al.}{2007}]{FRB_discov}
{Lorimer} D.~R.,  {Bailes} M.,  {McLaughlin} M.~A.,  {Narkevic} D.~J.,
  {Crawford} F.,  2007, \mn@doi [Science] {10.1126/science.1147532}, \href
  {http://adsabs.harvard.edu/abs/2007Sci...318..777L} {318, 777}

\bibitem[\protect\citeauthoryear{Lucarelli et~al.,}{Lucarelli
  et~al.}{2008}]{MAGIC_cpix}
Lucarelli F.,  et~al., 2008, \mn@doi [Nuclear Instruments and Methods in
  Physics Research] {http://dx.doi.org/10.1016/j.nima.2008.03.007}, 589, 415

\bibitem[\protect\citeauthoryear{{Lunnan} et~al.,}{{Lunnan}
  et~al.}{2014}]{lunnan2014}
{Lunnan} R.,  et~al., 2014, \mn@doi [\apj] {10.1088/0004-637X/787/2/138}, \href
  {http://adsabs.harvard.edu/abs/2014ApJ...787..138L} {787, 138}

\bibitem[\protect\citeauthoryear{{Lyne} \& {Graham-Smith}}{{Lyne} \&
  {Graham-Smith}}{2005}]{Lyne2005}
{Lyne} A.~G.,  {Graham-Smith} F.,  2005, {Pulsar Astronomy}

\bibitem[\protect\citeauthoryear{{Lyubarsky}}{{Lyubarsky}}{2014}]{lyubarsky2014}
{Lyubarsky} Y.,  2014, \mn@doi [\mnras] {10.1093/mnrasl/slu046}, \href
  {http://adsabs.harvard.edu/abs/2014MNRAS.442L...9L} {442, L9}

\bibitem[\protect\citeauthoryear{{Lyutikov}}{{Lyutikov}}{2002}]{lyutikov2002}
{Lyutikov} M.,  2002, \mn@doi [\apjl] {10.1086/345493}, \href
  {http://adsabs.harvard.edu/abs/2002ApJ...580L..65L} {580, L65}

\bibitem[\protect\citeauthoryear{{Lyutikov}, {Burzawa}  \& {Popov}}{{Lyutikov}
  et~al.}{2016}]{Lyutikov2016}
{Lyutikov} M.,  {Burzawa} L.,   {Popov} S.~B.,  2016, \mn@doi [\mnras]
  {10.1093/mnras/stw1669}, \href
  {http://adsabs.harvard.edu/abs/2016MNRAS.462..941L} {462, 941}

\bibitem[\protect\citeauthoryear{{Marcote} et~al.,}{{Marcote}
  et~al.}{2017}]{marcote2017}
{Marcote} B.,  et~al., 2017, \mn@doi [\apjl] {10.3847/2041-8213/834/2/L8},
  \href {http://adsabs.harvard.edu/abs/2017ApJ...834L...8M} {834, L8}

\bibitem[\protect\citeauthoryear{{Meyer}, {Horns}  \& {Zechlin}}{{Meyer}
  et~al.}{2010}]{meyer2010}
{Meyer} M.,  {Horns} D.,   {Zechlin} H.-S.,  2010, \mn@doi [\aap]
  {10.1051/0004-6361/201014108}, \href
  {http://adsabs.harvard.edu/abs/2010A%26A...523A...2M} {523, A2}

\bibitem[\protect\citeauthoryear{{Michilli} et~al.,}{{Michilli}
  et~al.}{2018}]{Michilli2018}
{Michilli} D.,  et~al., 2018, \mn@doi [\nat] {10.1038/nature25149}, \href
  {http://adsabs.harvard.edu/abs/2018Natur.553..182M} {553, 182}

\bibitem[\protect\citeauthoryear{{Murase}, {Kashiyama}  \&
  {M{\'e}sz{\'a}ros}}{{Murase} et~al.}{2016}]{Murase2016}
{Murase} K.,  {Kashiyama} K.,   {M{\'e}sz{\'a}ros} P.,  2016, \mn@doi [\mnras]
  {10.1093/mnras/stw1328}, \href
  {http://adsabs.harvard.edu/abs/2016MNRAS.461.1498M} {461, 1498}

\bibitem[\protect\citeauthoryear{{Patrignani} \& {Particle Data
  Group}}{{Patrignani} \& {Particle Data Group}}{2016}]{pdg_2016}
{Patrignani} C.,  {Particle Data Group} 2016, \mn@doi [Chinese Physics C]
  {10.1088/1674-1137/40/10/100001}, \href
  {http://cdsads.u-strasbg.fr/abs/2016ChPhC..40j0001P} {40, 100001}

\bibitem[\protect\citeauthoryear{{Pen} \& {Connor}}{{Pen} \&
  {Connor}}{2015}]{pen2015}
{Pen} U.-L.,  {Connor} L.,  2015, \mn@doi [\apj] {10.1088/0004-637X/807/2/179},
  \href {http://adsabs.harvard.edu/abs/2015ApJ...807..179P} {807, 179}

\bibitem[\protect\citeauthoryear{{Petroff} et~al.,}{{Petroff}
  et~al.}{2016}]{petroff2016}
{Petroff} E.,  et~al., 2016, \mn@doi [\pasa] {10.1017/pasa.2016.35}, \href
  {http://adsabs.harvard.edu/abs/2016PASA...33...45P} {33, e045}

\bibitem[\protect\citeauthoryear{{Popov} \& {Postnov}}{{Popov} \&
  {Postnov}}{2013}]{Popov2013}
{Popov} S.~B.,  {Postnov} K.~A.,  2013, preprint, \href
  {http://adsabs.harvard.edu/abs/2013arXiv1307.4924P} {} (\mn@eprint {arXiv}
  {1307.4924})

\bibitem[\protect\citeauthoryear{{Rane} \& {Lorimer}}{{Rane} \&
  {Lorimer}}{2017}]{Rane2017}
{Rane} A.,  {Lorimer} D.,  2017, \mn@doi [Journal of Astrophysics and
  Astronomy] {10.1007/s12036-017-9478-1}, \href
  {http://adsabs.harvard.edu/abs/2017JApA...38...55R} {38, 55}

\bibitem[\protect\citeauthoryear{{Rolke}, {L{\'o}pez}  \& {Conrad}}{{Rolke}
  et~al.}{2005}]{Rolke}
{Rolke} W.~A.,  {L{\'o}pez} A.~M.,   {Conrad} J.,  2005, \mn@doi [Nuclear
  Instruments and Methods in Physics Research A] {10.1016/j.nima.2005.05.068},
  \href {http://adsabs.harvard.edu/abs/2005NIMPA.551..493R} {551, 493}

\bibitem[\protect\citeauthoryear{{Scholz} et~al.,}{{Scholz}
  et~al.}{2016}]{scholz2016}
{Scholz} P.,  et~al., 2016, in AAS/High Energy Astrophysics Division. p. 105.03

\bibitem[\protect\citeauthoryear{{Scholz} et~al.,}{{Scholz}
  et~al.}{2017}]{scholz2017}
{Scholz} P.,  et~al., 2017, \mn@doi [\apj] {10.3847/1538-4357/aa8456}, \href
  {http://adsabs.harvard.edu/abs/2017ApJ...846...80S} {846, 80}

\bibitem[\protect\citeauthoryear{{Shearer}, {Stappers}, {O'Connor}, {Golden},
  {Strom}, {Redfern}  \& {Ryan}}{{Shearer} et~al.}{2003}]{Shearer2003}
{Shearer} A.,  {Stappers} B.,  {O'Connor} P.,  {Golden} A.,  {Strom} R.,
  {Redfern} M.,   {Ryan} O.,  2003, \mn@doi [Science]
  {10.1126/science.1084919}, \href
  {http://adsabs.harvard.edu/abs/2003Sci...301..493S} {301, 493}

\bibitem[\protect\citeauthoryear{{Spitler} et~al.,}{{Spitler}
  et~al.}{2014}]{spitler2014}
{Spitler} L.~G.,  et~al., 2014, \mn@doi [\apj] {10.1088/0004-637X/790/2/101},
  \href {http://adsabs.harvard.edu/abs/2014ApJ...790..101S} {790, 101}

\bibitem[\protect\citeauthoryear{{Spitler} et~al.,}{{Spitler}
  et~al.}{2016}]{FRB_repeater}
{Spitler} L.~G.,  et~al., 2016, \mn@doi [Nature] {10.1038/nature17168}, \href
  {http://adsabs.harvard.edu/abs/2016Natur.531..202S} {531, 202}

\bibitem[\protect\citeauthoryear{{Strader} et~al.,}{{Strader}
  et~al.}{2013}]{Strader2013}
{Strader} M.~J.,  et~al., 2013, \mn@doi [\apjl] {10.1088/2041-8205/779/1/L12},
  \href {http://adsabs.harvard.edu/abs/2013ApJ...779L..12S} {779, L12}

\bibitem[\protect\citeauthoryear{{Tendulkar}, {Kaspi}  \& {Patel}}{{Tendulkar}
  et~al.}{2016}]{tendulkar2016}
{Tendulkar} S.~P.,  {Kaspi} V.~M.,   {Patel} C.,  2016, \mn@doi [\apj]
  {10.3847/0004-637X/827/1/59}, \href
  {https://ui.adsabs.harvard.edu/#abs/2016ApJ...827...59T} {827, 59}

\bibitem[\protect\citeauthoryear{{Tendulkar} et~al.,}{{Tendulkar}
  et~al.}{2017}]{tendulkar2017}
{Tendulkar} S.~P.,  et~al., 2017, \mn@doi [\apjl] {10.3847/2041-8213/834/2/L7},
  \href {http://adsabs.harvard.edu/abs/2017ApJ...834L...7T} {834, L7}

\bibitem[\protect\citeauthoryear{{Thornton} et~al.,}{{Thornton}
  et~al.}{2013}]{Thornton2013}
{Thornton} D.,  et~al., 2013, \mn@doi [Science] {10.1126/science.1236789},
  \href {http://adsabs.harvard.edu/abs/2013Sci...341...53T} {341, 53}

\bibitem[\protect\citeauthoryear{{Vieyro}, {Romero}, {Bosch-Ramon}, {Marcote}
  \& {del Valle}}{{Vieyro} et~al.}{2017}]{vieyro2017}
{Vieyro} F.~L.,  {Romero} G.~E.,  {Bosch-Ramon} V.,  {Marcote} B.,   {del
  Valle} M.~V.,  2017, \mn@doi [\aap] {10.1051/0004-6361/201730556}, \href
  {http://adsabs.harvard.edu/abs/2017A%26A...602A..64V} {602, A64}

\bibitem[\protect\citeauthoryear{{Waxman}}{{Waxman}}{2017}]{Waxman2017}
{Waxman} E.,  2017, \mn@doi [\apj] {10.3847/1538-4357/aa713e}, \href
  {http://adsabs.harvard.edu/abs/2017ApJ...842...34W} {842, 34}

\bibitem[\protect\citeauthoryear{{Zanin} et~al.}{{Zanin} et~al.}{2013}]{MARS}
{Zanin} R.,  et~al., 2013, in {Proceedings, 33rd International Cosmic Ray
  Conference (ICRC2013): Rio de Janeiro, Brazil, July 2-9, 2013}. p.~0773, \url
  {http://www.cbpf.br/%7Eicrc2013/papers/icrc2013-0773.pdf}

\bibitem[\protect\citeauthoryear{{Zhang}}{{Zhang}}{2018}]{zhang2018}
{Zhang} B.,  2018, \mn@doi [\apjl] {10.3847/2041-8213/aaadba}, \href
  {http://adsabs.harvard.edu/abs/2018ApJ...854L..21Z} {854, L21}

\bibitem[\protect\citeauthoryear{{Zhang} \& {Zhang}}{{Zhang} \&
  {Zhang}}{2017}]{zhang2017}
{Zhang} B.-B.,  {Zhang} B.,  2017, \mn@doi [\apjl] {10.3847/2041-8213/aa7633},
  \href {http://adsabs.harvard.edu/abs/2017ApJ...843L..13Z} {843, L13}

\makeatother
\end{thebibliography}

\section*{Affiliations}
\begin{flushleft}
$^{1}$ {Inst. de Astrof\'isica de Canarias, E-38200 La Laguna, and Universidad de La Laguna, Dpto. Astrof\'isica, E-38206 La Laguna, Tenerife, Spain} \\
$^{2}$ {Universit\`a di Udine, and INFN Trieste, I-33100 Udine, Italy} \\
$^{3}$ {National Institute for Astrophysics (INAF), I-00136 Rome, Italy} \\
$^{4}$ {ETH Zurich, CH-8093 Zurich, Switzerland} \\
$^{5}$ {Universit\`a di Padova and INFN, I-35131 Padova, Italy} \\
$^{6}$ {Technische Universit\"at Dortmund, D-44221 Dortmund, Germany} \\
$^{7}$ {Croatian MAGIC Consortium: University of Rijeka, 51000 Rijeka, University of Split - FESB, 21000 Split,  University of Zagreb - FER, 10000 Zagreb, University of Osijek, 31000 Osijek and Rudjer Boskovic Institute, 10000 Zagreb, Croatia.} \\
$^{8}$ {Saha Institute of Nuclear Physics, HBNI, 1/AF Bidhannagar, Salt Lake, Sector-1, Kolkata 700064, India} \\
$^{9}$ {Max-Planck-Institut f\"ur Physik, D-80805 M\"unchen, Germany} \\
$^{10}$ {now at Centro Brasileiro de Pesquisas F\'isicas (CBPF), 22290-180 URCA, Rio de Janeiro (RJ), Brasil} \\
$^{11}$ {Unidad de Part\'iculas y Cosmolog\'ia (UPARCOS), Universidad Complutense, E-28040 Madrid, Spain} \\
$^{12}$ {University of \L\'od\'z, Department of Astrophysics, PL-90236 \L\'od\'z, Poland} \\
$^{13}$ {Deutsches Elektronen-Synchrotron (DESY), D-15738 Zeuthen, Germany} \\
$^{14}$ {Institut de F\'isica d'Altes Energies (IFAE), The Barcelona Institute of Science and Technology (BIST), E-08193 Bellaterra (Barcelona), Spain} \\
$^{15}$ {Universit\`a  di Siena and INFN Pisa, I-53100 Siena, Italy} \\
$^{16}$ {Universit\`a di Pisa, and INFN Pisa, I-56126 Pisa, Italy} \\
$^{17}$ {Universit\"at W\"urzburg, D-97074 W\"urzburg, Germany} \\
$^{18}$ {Finnish MAGIC Consortium: Tuorla Observatory and Finnish Centre of Astronomy with ESO (FINCA), University of Turku, Vaisalantie 20, FI-21500 Piikki\"o, Astronomy Division, University of Oulu, FIN-90014 University of Oulu, Finland} \\
$^{19}$ {Departament de F\'isica, and CERES-IEEC, Universitat Aut\'onoma de Barcelona, E-08193 Bellaterra, Spain} \\
$^{20}$ {Japanese MAGIC Consortium: ICRR, The University of Tokyo, 277-8582 Chiba, Japan; Department of Physics, Kyoto University, 606-8502 Kyoto, Japan; Tokai University, 259-1292 Kanagawa, Japan; RIKEN, 351-0198 Saitama, Japan} \\
$^{21}$ {Inst. for Nucl. Research and Nucl. Energy, Bulgarian Academy of Sciences, BG-1784 Sofia, Bulgaria} \\
$^{22}$ {Universitat de Barcelona, ICC, IEEC-UB, E-08028 Barcelona, Spain} \\
$^{23}$ {Humboldt University of Berlin, Institut f\"ur Physik D-12489 Berlin Germany} \\
$^{24}$ {also at Dipartimento di Fisica, Universit\`a di Trieste, I-34127 Trieste, Italy}\\
$^{25}$ {also at Port d'Informaci\'o Cient\'ifica (PIC) E-08193 Bellaterra (Barcelona) Spain} \\
$^{26}$ {also at INAF-Trieste and Dept. of Physics \& Astronomy, University of Bologna}\\
$^{27}$Joint Institute for VLBI ERIC, Postbus 2, 7990 AA Dwingeloo, The Netherlands\\
$^{28}$Max-Planck-Institut f\"ur Radioastronomie, Auf dem H\"ugel 69, Bonn, D-53121, Germany\\
$^{29}$ASTRON, Netherlands Institute for Radio Astronomy, Postbus 2, 7990 AA, Dwingeloo, The Netherlands\\
$^{30}$Anton Pannekoek Institute for Astronomy, University of Amsterdam, Science Park 904, 1098 XH Amsterdam, The Netherlands\\
$^{31}$Department of Physics, the University of Tokyo, Bunkyo, Tokyo 113-0033, Japan\\
$^{32}$Department of Physics, the Pennsylvania State University, University Park, PA 16802, USA\\
$^{33}$National Astronomy and Ionosphere Center, Arecibo Observatory, Puerto Rico, 00612, USA \\
\end{flushleft}







\bsp	
\label{lastpage}
\end{document}